\documentclass[prd,nofootinbib,twocolumn]{revtex4}
\usepackage{amsmath}
\usepackage{amssymb}
\usepackage{graphicx}
\usepackage[usenames, dvipsnames]{color}

\newcommand{\blue}{\textcolor{blue}}

\usepackage{graphicx}

\newcommand{\Lk}{L_{\pm}^{\rm K  }}

\newcommand{\code}{\texttt}

\usepackage{amsmath}
\usepackage{graphicx}
\usepackage[colorlinks=true, allcolors=blue]{hyperref}

\begin{document}
\title{Angular Momentum for Black Hole Binaries in Numerical Relativity} 
\author{Ritesh Bachhar}
\affiliation{Department of Physics, East Hall, University of Rhode Island, Kingston, RI 02881}
\author{Richard H.~Price}
\affiliation{Department of Physics, MIT, 77 Massachusetts Ave., Cambridge, MA 02139}
\author{Gaurav Khanna}
\affiliation{Department of Physics and Center for Computational Research, University of Rhode Island, Kingston, RI 02881}
\affiliation{Department of Physics and Center for Scientific Computing \& Data Research, University of Massachusetts, Dartmouth, MA 02747}


\begin{abstract}
\noindent The extensive catalog of waveforms, with details of binary black hole inspiral and merger, offer an opportunity to understand black hole interactions beyond the large separation regime. We envision a research program that focuses on the transfer of angular momentum from spin of the individual holes to the orbital angular momentum and the role of tidal coupling in the process. 
That analysis will require the formulation of an expression for the orbital angular momentum of a binary, an expression that is useful at small separations, since that regime is well out of the range of Newtonian approximations and is where tidal coupling should be most interesting.  
 We report here such an expression, a binary orbital angular momentum based on numerical relativity results for quasi-circular orbits, that agrees remarkably well with a similar quantity constructed with particle-perturbation techniques for the Kerr geometry.


\end{abstract}

\maketitle

\section{Introduction}
Computational modeling  of binary black hole dynamics, once a fantasy,
is now so mature that a publicly accessible summary of results for more 
than two thousand models, i.e., the SXS catalog,~\cite{SXScatalog,SXSpaper} 
is available. These models start at separations large
enough that they can be well approximated by a model of two point
masses, moving according to Newtonian physics, and losing energy and
angular momentum in gravitational wave (GW) emission approximated by the
quadrupole formula. This relatively simple approach, and its extensions, have  been known as
the particle-perturbation method (PP), and were the main tool for
understanding binary inspiral until the conquest by numerical
relativity (NR)~\cite{NR} in the early part of the new millennium.

Numerical relativity can do what PP cannot do.  Highly developed
numerical relativity (NR) codes extend binary black hole evolution from the
early PP-appropriate era to the formation and ringdown of the final
remnant black hole.  In principle, these numerical evolutions give us
a laboratory for extracting nonlinear effects through comparisons using
PP and with insights and vocabulary based on the PP picture.

Our primary interest in exploiting the NR results is the role of black
hole spin.  The SXS catalog includes the spins of each of the two initial 
black holes, and the spin of the single remnant hole. Intuition
suggests that negligible angular momentum can be radiated during the
plunge following the binary equivalent~\cite{binaryISCO} of the innermost 
stable circular orbit (ISCO). It is easily verified that  the angular momentum in the remnant 
Kerr hole is well approximated by the binary orbital angular momentum 
(BOAM) at the plunge~\cite{BlandfordHughes}.
For cases in which the initial holes have spin, most of that spin
angular momentum appears in the Kerr remnant as an addition to the
ISCO orbital angular momentum~\cite{BuoKidderLehner}. 

It is important to note that for the inspiral of holes with initial spins, the role of spin is not
simply a storage of angular momentum until the remnant is formed. This
can be seen in the effect of the spins on the evolution of the
binary. Cases in which there is significant spin aligned in the same
direction as the BOAM evolve more slowly than spinless evolution;
cases with initial spins anti-aligned, evolve more quickly.
We take this as a hint that the angular momentum of the spins is being
transferred to the BOAM. In the case of aligned BOAM-spin this transfer
increases the BOAM requiring increased time for the GW radiation of
the BOAM, and for the inward evolution of the binary. The opposite is the case for initial anti-alignment.

This transfer is what is usually called spin-orbit coupling, and --
for ordinary astrophysical bodies -- is understood as brought about 
by tidal torque. That torque is due to the tidal bulge\footnote{Here the word 
``bulge" is meant to be suggestive only. Since the shape of the horizon can be given different meanings, we will not ascribe quantitative meaning to the ``bulge," or to the tidal Love number related to that bulge.} 
on the torqued body being out of alignment with the line connecting the center of the 
torqued object and the source of the tidal effects. The root of this 
misalignment is the viscosity of the material in the torqued object.

Black holes have no ``material.'' By black hole we typically mean an
event horizon, a ``teleologically'' defined mathematical surface, a
surface defined by the no-escape condition in the infinite future.  How such a
non-material entity can have viscosity is therefore of great interest.
Work on this question dates back at least to 1973~\cite{hartle,hartle2,price-whelan}.
That work was chiefly based on the effect of a point particle perturbation 
on a Kerr hole~\cite{poisson}. More recently, Poisson~\cite{poisson2} has compared black
hole process and those of fluid bodies. That work should give us at least a very approximate measure of how much angular momentum is transferred via tidal coupling. The results presented here will be a necessary first step in determining whether the angular momentum of the binary remnant is compatible with the limits suggested by those approximations.
In brief, the strategy we will follow is to start with NR data, available from a ``hybridized'' 
surrogate model NRHybSur3dq8~\cite{hybrid} trained with NR and Post-Newtonian waveforms. 
In this paper, we will use that NR data for evolution of a binary with no initial spins. From that 
NR data we establish a BOAM that is useful in considerations of angular momentum conservation. 
It is interesting  that we find that this NR-based BOAM agrees better than might be expected with BOAM 
inferred from the PP results for the Kerr geometry.

The NR-based BOAM will be used in a subsequent paper~\cite{PaperII} to infer the evolution of the 
black hole spins from conservation of angular momentum. That spin evolution will then be compared to 
simple models of tidal torque based on the strength of tidal effects and difference of the orbital 
velocity of a hole and the angular velocity $\Omega_H$~\cite{poisson2} of its event horizon. 




The rest of this paper is organized as follows. In Sec.~\ref{sub:SXS} we briefly review 
the SXS data relevant to the role of spin in binary evolution, then in Sec.~\ref{sub:Lidea}  
we show how the NR data for GW emission can be used to construct a BOAM that is meaningful for a 
study of spin-orbit exchange of angular momentum. In Sec.~\ref{sec:WhatL} we present numerical results 
for the BOAM identified in Sec.~\ref{sec:Background}, compare it with the Newtonian approximation, and 
use a fitting to produce what will be necessary for our study of tidal interaction: the NR-based BOAM 
as a function of the orbital velocity. Section~\ref{sec:KerrInspired} presents our method for adapting, from PP theory, the angular momentum for a point particle orbiting a Kerr hole to a binary of comparable masses.
We summarize our conclusions in Sec.~\ref{sec:Conc}.

Throughout this	paper we generally use units in which $c=G=1$, although factors of $G$ 
are sometimes explicit. In	our plots, all	quantities will	be normalized by 
the total system mass $M$. We will consistently use the symbol $L$ to denote BOAM, 
while $J$ will mean angular momentum more generally, in particular in connection with GW emission. We 
will follow the practice of Ref.~\cite{SXSpaper} of using the symbol $\chi$ to denote the angular momentum of 
an individual hole divided by the square of the mass of that hole, and for positive $\chi$ to indicate 
alignment with the BOAM.

\begin{figure*}
    \centering
    \includegraphics[width=.95\textwidth]{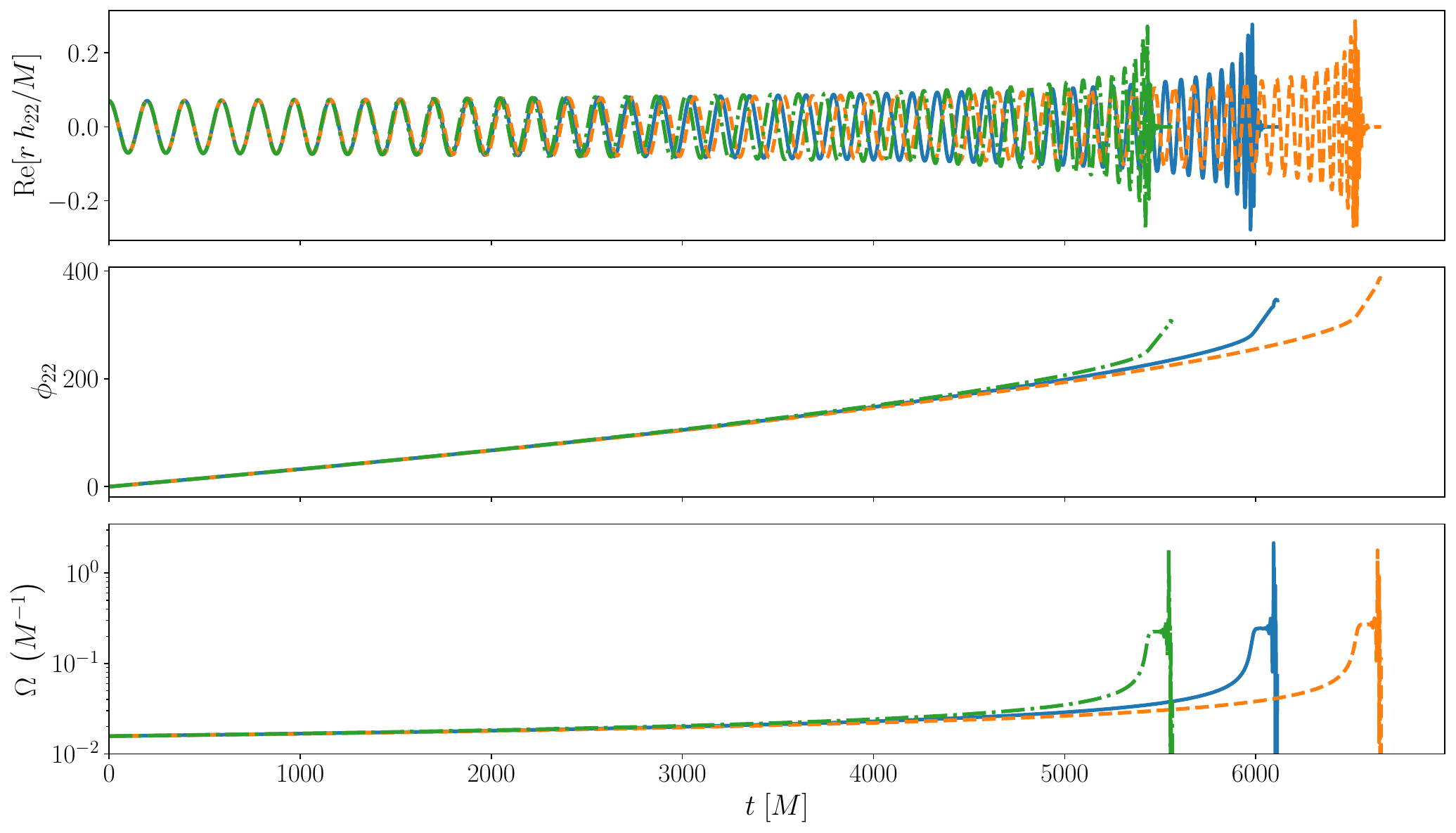}
    \caption{Real part of $(2,2)$ waveform (top panel), phase of $(2,2)$ mode (middle panel), and orbital frequency (bottom panel) as a function of time, for equal mass ratio binaries with no initial spins (blue, solid), with initial spins $\chi = +0.3$ (orange, dashed) for each, and for $\chi = -0.3$ (green, dash-dot) for each. Waveforms are aligned in time such that all three have $\Omega = 0.0157/M$ at the start of evolution. These waveforms correspond to the SXS catalog models BBH:0161 ($\chi=-0.3$), 0001 ($\chi=0$), and 0164 ($\chi=+0.3$). }
\label{fig:waveform}
\end{figure*}

We start, in Fig.~\ref{fig:waveform}, with a suggestive look at a possible  effect of 
angular momentum transfer. In that figure we see the phenomenon mentioned above. As 
compared to the waveform for no initial spins, the one for initial spins aligned with 
BOAM is slower to reach the end of quasicircular orbiting, and the waveform for the 
initially anti-aligned spins is faster. As expected, the shifts among waveform phases 
are dramatic as the end of quasicircular orbiting is approached. What is encouraging 
about these NR results, however, is that there is significant differences of phase 
among the three cases, sufficiently early that they can be understood with the approximation 
methods we will be using.

Although Fig.~\ref{fig:waveform} is an encouraging introduction, the GW waveforms 
are not the optimal data for understanding the details of black hole interactions. Throughout 
this paper, and subsequent investigations including spin, our go-to technique will be to look 
at various kinematic and dynamic quantities as functions of the orbital or gravitational 
wave angular frequency. It will become clear that this approach often simplifies comparisons, and avoids ambiguities. Note that throughout this paper, time $t$ refers to the Schwarzschild time coordinate, equivalently the time at infinity. 

Lastly, it is worth emphasizing that the interactions we are pursuing are of primary importance in the strong field regime. We are, in fact, using NR results to understand  the strong field regime. While well-established techniques, like the post-Newtonian (PN) approximations are of great importance for the early evolution of the binary, we do not make use of them here.

\section{Background: spin-orbit coupling evidence from NR and our approach}\label{sec:Background}
\subsection{Evidence from SXS}\label{sub:SXS}
The ISCO~\cite{binaryISCO} represents a sharp change in the evolution of the binary.
Prior to the ISCO the evolution is a slow quasi-adiabatic decrease in the radius of 
the circular binary orbit~\cite{circularization}. It 
is in the pre-ISCO motion that GWs carry away energy and angular momentum of the system. 
The plunge subsequent to the ISCO is rapid, and it is a reasonable assumption that it 
entails negligible loss of energy or angular momentum.

This suggests~\cite{BlandfordHughes,PriceKhanna} that, in the case of holes with no 
initial spin, the BOAM at the ISCO shows up as the spin of the ``remnant" Kerr hole 
that is the end point of the evolution. Buonanno {\it et al.}~\cite{BuoKidderLehner} have 
taken a step further, and have shown that in the case of holes with initial spins, 
the Kerr remant spin is well approximated by the vector sum of those spins and the BOAM 
at the ISCO.

While the appearance of orbital angular momentum in the spin of the remnant hole is 
intuitively appealing, the mechanism for the contribution of the individual hole spins 
is less obvious, and hence more interesting. Our working hypothesis is that the individual 
spins are transferred to the BOAM during the late pre-ISCO stage of inspiral. Evidence of 
this can be seen in SXS models with nonzero initial values of the dimensionless parameter 
$\chi$, the spin angular momentum of a hole divided by the square of the hole's mass. As an 
example we can compare equal mass models SXS:BBH:0161 and SXS:BBH:0164. In the former, 
each hole has spin $\chi=-0.3$, where the minus sign indicates anti-alignment of the spins 
with the BOAM. The latter model, SXS:BBH:0164, differs only in that initially it has 
aligned spin and BOAM, i.e., $\chi=+0.3$.~\cite{initialvsref} 

We note that for the $\chi=-0.3$ model, the time to the ISCO is approximately 4,250 in 
units of the total binary mass, while for the $+0.3$ model the time is 5,025. This fits 
our viewpoint that spin is being transferred to BOAM. 
For 0161, the $\chi=-0.3$ model, the transfer of angular momentum decreases the BOAM and 
decreases the time necessary to shed the BOAM as GW emission. The opposite applies to 0164, 
the $\chi=+0.3$ model.

The details of the GW waveforms, as proxies for the orbital angular motion, offer a somewhat 
deeper insight. The 0161 waveform phase starts to lead that of 0164 relatively early in the 
evolution (at time around 2,500) when the separation of the black holes is on the order of 14 
in units of the binary mass, a separation at which we might expect relativistic effects to appear 
but not dominate. At later times, the rate at which the phases of 0161 and 0164 deviate increases, 
as we would expect in view of the sensitivity of tidal effects to separation (the sixth power 
of the separation!~\cite{poisson,poisson2}).

It is important, at this point, to ask whether there is another explanation for what is observed 
in the comparison of the 0161 and 0164 SXS waveforms. Is there another explanation for the more 
rapid evolution of the binary with anti-aligned spins than with aligned spins. There are two such 
possibilities. (1) For a binary with spinning holes, the  binary energy and angular momentum are 
different than if the holes were nonspinning. Thus the time required for shedding energy and angular 
momentum in GWs would be affected. (2) The rate of GW emission of a binary is affected by the spins 
of the individual holes in the binary.

The first possibility deserves careful inspection since the ISCO for a point particle around a Kerr 
hole is different for a prograde and a retrograde orbit. For a Kerr hole with mass $M$ and spin parameter 
$a=0.3\,M$, the prograde/retrograde ISCOs are respectively at radii 4.979 and 6.949 in units of $M$. 
One would therefore not be surprised if the equivalent of the ISCO for a \emph{binary} would be smaller, 
and therefore later, for binaries with aligned spin than for anti-aligned spin.  The effect, a manifestation 
of GR effects in the interaction of the holes, might also be an element in the $\chi=0.3$ vs -0.3 phase 
shift noted in the earlier waveforms. This matter will be discussed in some depth in our subsequent paper~\cite{PaperII}. For now, we point out that it is a part of the motivation for the exploration of ideas in Sec.~\ref{sec:KerrInspired}.
The second alternative explanation of observed shifts, the sensitivity of GW emission to spins of the holes, 
seems unphysical, since the generation of the GW emission (at least for ``particles" moving well below 
lightspeed) is associated with the distant fields of the sources.

\subsection{A meaningful BOAM}\label{sub:Lidea}
We must ask just what it is that we want for a meaningful BOAM.  For our purposes, the total 
angular momentum (BOAM and spin) is lost to the system only in GW emission. In the case that 
the individual holes have no spin, we therefore seek a quantity that has this feature.

In the absence of spin effects, the BOAM $L$ that we want will therefore be the quantity that satisfies
\begin{equation}
    \frac{dL}{dt}=\dot{J}^{\rm GW}
\end{equation}
where $\dot{J}^{\rm GW}$ is the rate at which GW carries off angular momentum. 

If the approximation is to remain valid that the binary orbit remains nearly circular during inspiral, the angular momentum can safely be assumed to be a function of angular velocity ($\Omega$). Now our useful BOAM must satisfy,


\begin{equation}\label{eq:Ldot}
    \frac{dL}{dt}= \dot{\Omega}\frac{dL}{d\Omega},
\end{equation}
where the overdot represents a derivative with respect to time.
From Eq.~\eqref{eq:Ldot}, we have that $L(\Omega)$ is to be found from
\begin{equation}\label{eq:dLdOm}
    \frac{dL}{d\Omega}=\frac{ \dot{J}^{\rm GW}}{\dot{\Omega}}\,,
\end{equation}
in which all quantities on the right can be 
directly calculated from NR waveform \cite{surrkick}.

\section{Characterizing binary orbital angular momentum}\label{sec:WhatL}
\subsection{Newtonian form}
From Newtonian physics, for masses $m_1$
and $m_2$, at distances $r_1$, and $r_2$, from the barycenter,
the orbital     angular momentum is
\begin{equation}\label{eq:NewtBoam}
  L=L_1+L_2=m_1r_1^2\Omega+m_2r_2^2\Omega=M(GM)^{2/3}\Omega^{-1/3}\,f(q)
\end{equation}
where
\begin{equation}
  f(q)=\frac{1}{(1+q)(1+1/q)}\,.
\end{equation}

Here, and throughout this paper, we will find it interesting to
compare PP-based quantities with those from NR.  From the standard PP
result for quadrupole radiation~\cite{peters-mathews-MTW}
 we have
\begin{equation}
  \dot{E}^{\rm GW}=\frac{32}{5}\frac{M G^4m_1^2m_2^2}{c^5r^5}=\frac{32}{5}\frac{M (GM)^4}{c^5r^5}\,[f(q)]^2
\end{equation}
and the Kepler law
\begin{equation}
  \Omega=\sqrt{{GM}/{r^3}\;},
\end{equation}
where $r=r_1+r_2$ is the separation of the point masses. As a function of $\Omega$, the GW energy flux is
\begin{equation}\label{eq:Edot}
  \dot{E}^{\rm GW}=\frac{32}{5}\frac{M}{c^5}(GM)^{7/3}\Omega^{10/3}[f(q)]^2\,.
\end{equation}

In the Newtonian context, the rate of loss of angular momentum is related to $\dot{E}^{\rm GW}$, the rate of loss of energy due to GW emission by $\dot{E}^{\rm GW} =\Omega\dot{J}^{\rm GW}$. From Eqs.~\eqref{eq:NewtBoam}, \eqref{eq:Edot} and $dL/dt=\dot{J}^{\rm GW}$ this gives us
\begin{equation}\label{eq:OmDot}
\dot{\Omega}=  -\frac{96}{5c^5}(GM)^{5/3}\Omega^{11/3}f(q)\,.
\end{equation}
and hence 
\begin{equation}\label{eq:dLdOmNewt}
  \frac{\dot{J}^{\rm GW}}{\dot{\Omega}}=\frac{\dot{E}^{\rm GW}}{\dot{\Omega}\Omega}=-\frac{1}{3} M (GM)^{2/3}\Omega^{-4/3}f(q)\,.
\end{equation}
The right-hand side is the Newtonian form we want to compare to the NR results for the left-hand side.

\subsection{Results for numerically inspired BOAM}

To obtain the numerical data for the study of binary orbital parameters, we used a ``hybridized'' 
NR surrogate model \code{NRHybSur3dq8}. This model is trained with Post-Newtonian and EOB waveforms in the early inspiral and smoothly connects to NR data before the merger. Because of this, the model is capable of generating long waveforms, which are ideal for our investigation. 
We decomposed the complex dominant quadrupole mode ($l=|m|=2$), 
\begin{equation}
    h_{22} = A_{22}\,e^{-i\phi_{22}}
\end{equation}
into amplitude, $A_{22}$ and phase, $\phi_{22}$. The time derivative of the phase ($\phi_{22}$) 
gives us the angular frequency of the signal. Since the binary is a quasi-circular 
orbit we can calculate the orbital angular velocity ($\Omega$) of the binary using,
\begin{align}
    \omega = \frac{d\phi_{22}}{dt} \\
    \Omega = \frac{\omega}{2}.
\end{align}
Similarly, the time derivative of the orbital angular velocity ($\dot{\Omega}$) can be obtained 
from
\begin{align}
    \dot{\omega} = \frac{d^2\phi_{22}}{dt^2} \\
    \dot{\Omega} = \frac{\dot{\omega}}{2}.
\end{align}

\begin{figure*}
    \centering
    \includegraphics[width=0.95\textwidth]{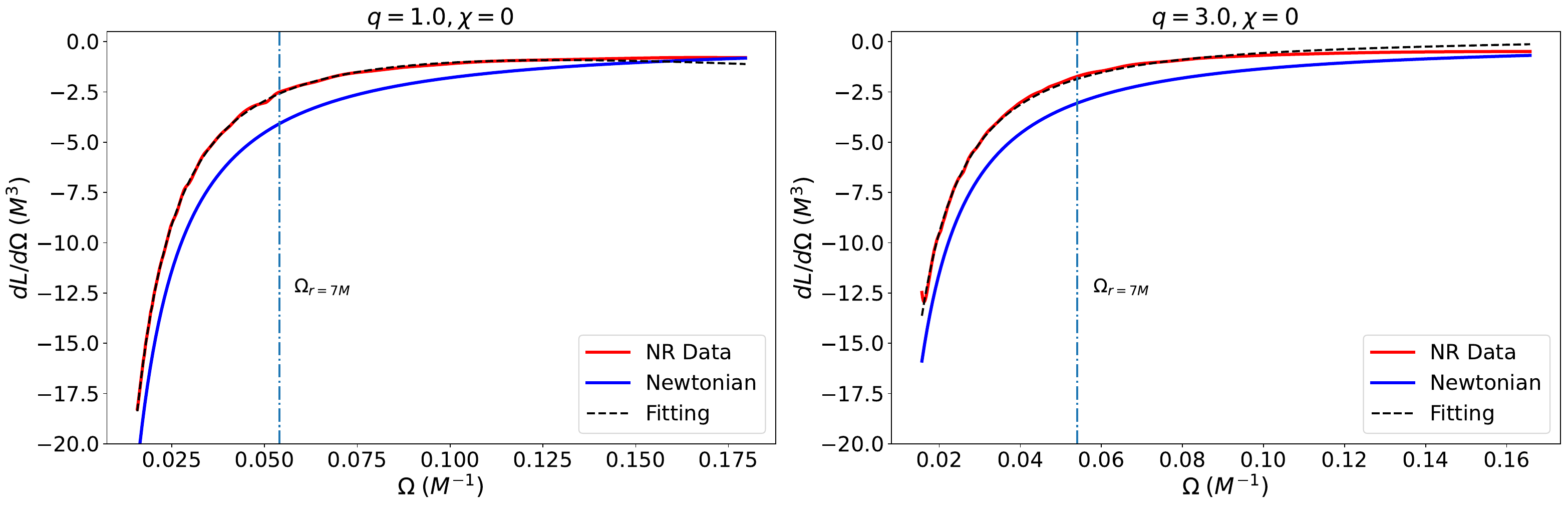}
    \caption{$dL/d\Omega$ as a function of angular velocity $\Omega$ for $q=1, 3$ with no initial spin. 
    NR data is compared with the Newtonian form and a simple fit is provided. The vertical line corresponds to $dL/d\Omega$ at angular velocity $\Omega=\sqrt{GM/r^3\;}$ for $r=7M$, which approximately coincides with the ISCO. The dashed curves show the NR-based fitting given by Eqs.~\eqref{eq:dLdOmega}. 
    }
\label{fig:dLdOm}
\end{figure*}

\begin{figure*}
    \centering
    \includegraphics[width=0.95\textwidth]{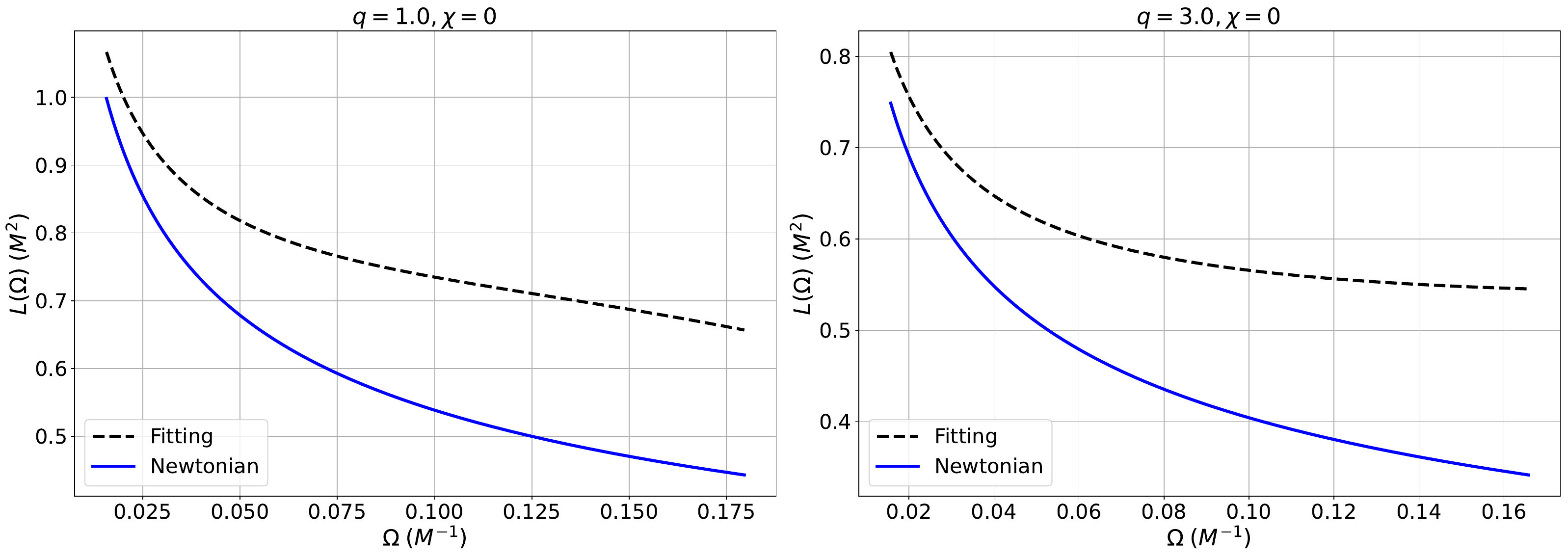}
    \caption{Angular momentum ($L$) as a function of angular velocity ($\Omega$) for mass ratio $q=1$ (left) and  $q=3$ (right) with no initial spin. The Newtonian angular momentum is represented as the blue solid line, while the black dashed line depicts Eqs.~\eqref{eq:LOmega1} for $q=1$, and \eqref{eq:LOmega2} for $q=3$. The Newtonian, and fitting curves asymptotically approach the same values as $\Omega$ becomes small.}
\label{fig:LOmega}
\end{figure*}

We adapted the method explained in Ref.~\cite{surrkick} to calculate the angular momentum radiated by gravitational waves ($\dot{J}^{\rm GW}$). Now we have everything to calculate the 
left-hand side of Eq.~\eqref{eq:dLdOmNewt}. In Fig.~\ref{fig:dLdOm} we have compared the result obtained from NR data to the Newtonian result. To find a meaningful BOAM for the study of the transfer of angular momentum via tidal coupling, we provide empirical fits for Eq.~\eqref{eq:dLdOm} as a function of orbital frequency, for $q=1$ and $q=3$ without spin,

\begin{equation}\label{eq:dLdOmega}
    \begin{aligned}
    \frac{dL}{d\Omega}\bigg|_{q=1} &= -\bigg(\frac{1}{12} - 0.746\Omega + 2.612 \Omega^{2} + 13.721 \Omega^{3}\bigg) \Omega^{-4/3}\;, \\
    \frac{dL}{d\Omega}\bigg|_{q=3} &= -\bigg(\frac{1}{16} -0.618 \Omega + 3.577 \Omega^{2} -10.178 \Omega^{3}\bigg) \Omega^{-4/3}
    \end{aligned}
    \end{equation}
The choice of these functions is made such that they converge to Newtonian results in a low orbital frequency limit. $dL/d\Omega$ as a function of $\Omega$ is presented in Fig.~\ref{fig:dLdOm}.


\section{Kerr-inspired BOAM}\label{sec:KerrInspired}

Our development of an appropriate expression for the BOAM starts by
adapting the particle motion around a Kerr hole as presented, for 
example, in Refs.~\cite{Bini, bardeen}. 
In doing so we are omitting the effect of spin curvature of the smaller body in our calculation, which in fact has negligible effect on the circular equatorial orbit~\cite{motionInKerr}. 
We present the expressions from Ref.~\cite{bardeen}
for the conserved angular momentum of a particle of mass $\mu$ in a circular orbit of radius $r$ around a Kerr hole of mass $M$. The meaning of the $\pm$ and $\mp$ signs is related to prograde and retrograde particle motion, and is explained in Ref.~\cite{bardeen}.
\begin{equation}\label{eq:Lk}
  \frac{\Lk}{\mu}=\frac{\pm M^{1/2}\left(r^2 \mp 2aM^{1/2}r^{1/2} + a^2\right)}{r^{3/4}\left(r^{3/2} - 3Mr^{1/2}\pm 2aM^{1/2}\right)^{1/2}}
\end{equation}
To treat hole 2 as the ``particle" we set $\mu=m_2$, $M=m_1$, $a=\chi_1m_1$, and take $r$ to be the separation of the holes. Here and below, treating hole 1 as the particle only requires exchanging 1 and 2 on appropriate quantities. 

The result of this procedure is {\it not} what we want since it gives $L^{\rm K}_{end}$, the angular velocity 
about one of the black holes, as a center of rotation. We must adjust this to find the physical angular velocity 
$L^{\rm K}_b$, the angular velocity about the barycenter.  To do this we multiply $L^{\rm K}_{end}$ by the factor that converts 
$L^{\rm K}_{end}$ to $L^{\rm K}_b$ in the Newtonian case. Our notations here: $M=m_1+m_2$ is the total mass; 
$r_1$ and $r_2$ are distances from the binary's barycenter to the individual holes, with $r_1=r(m_2/M),\; r_2=r(m_1/M)$, where $r$ is the distance between them. 
The reasons behind choosing this approach of calculating BOAM are twofold. First, our method treats each hole on an equal footing, i.e., the spin of each hole is considered in the BOAM calculation. Second, we have discovered in preparation for the following paper ~\cite{PaperII} that the process mentioned in this section estimates the final BH spin better than other theoretically motivated methods such as~\cite{BuoKidderLehner, PriceKhanna} over extended parameter space.

Below, we use ``end'' to refer to  dynamical quantities based on one of the holes ($m_1$ or $m_2$)
as stationary inertial points and ``b'' as  quantities calculated with the barycenter as the stationary
inertial point. 
%

We find that the Newtonian  angular momentum referred to the two different centers are
\begin{equation}
L_{e1}=m_1\sqrt{Gm_2\;}r^{1/2}\ \mbox{and}\ L_{b1}=m_1\sqrt{Gm_2\;}r_1^{3/2}r^{-1}.
\end{equation}
For the Newtonian expressions then, the conversion from the end value to the barycenter value for hole 1 is
 \begin{equation}\label{eq:Gamma1}
    \Gamma_1\equiv\frac{L_{b1}}{L_{e1}}=(1+q)^{-3/2}.
\end{equation} 
The analogous multiplier for particle 2 follows by substituting $1/q$ for $q$.


We now use 
\blue{Eq.~\eqref{eq:Lk}} to find the Kerr-inspired angular momentum of hole 1 about hole 2, then convert it to the angular momentum about the barycenter 
using the same conversion factor $\Gamma_1$, as in Eq.~\eqref{eq:Gamma1}. The same procedure is used for the motion of hole 2. 

Finally, with 
$L_1^{end}$ and $L_2^{end}$ from Eq.~\eqref{eq:Lk}, for  
 appropriate choices of radii and masses, we take the BOAM, $ L_{\rm bin}^{\rm K} $, to be $\Gamma_1L_1^{end}+\Gamma_2L_2^{end}$. 
 The result is
\begin{displaymath}
     L_{\rm bin}^{\rm K}=\frac{\sqrt{GMr\;}M}{(1+q)(1+1/q)}
\end{displaymath}
\begin{equation}
    \times\left[
     \frac{1}{\sqrt{\left(1+q\right)\left(1+q-\frac{3M}{r}\right)\;}}+ \frac{1}{\sqrt{\left(1+q^{-1}\right)\left(1+q^{-1}-\frac{3M}{r}\right)\;}}
    \right]
\end{equation} 
when both black holes don't have spin. The Newtonian equivalent differs only in the absence of the $3M/r$ subtractions inside the square roots.  
The results in Tables~\ref{tab:Lbinq1nospin} - \ref{tab:Lbinq3spin} 
show that the Kerr-inspired values  differ significantly from the 
Newtonian values only when the separation is small and one would expect significant general relativistic effects. 
 
What is most important to the purpose of this paper  is to compare these Kerr-inspired results for the BOAM 
with a measure of the BOAM that has meaning for conservation. This is the NR content of the right-hand side 
of Eq.~\eqref{eq:dLdOm} in which the meaningful BOAM is taken to be the quantity that is decreased by the 
angular momentum in outgoing GWs. 

Integrating the fits given in Eq.~\eqref{eq:dLdOmega}, we arrive at the expressions,
\begin{equation}\label{eq:LOmega1}
    L_{\rm bin}^{\rm GW}\bigg|_{\rm q = 1}= \left(\frac{1}{4} +1.119\Omega -1.567\Omega^2 -5.145\Omega^3 \right)\Omega^{-1/3}
\end{equation}
\begin{equation}\label{eq:LOmega2}
    L_{\rm bin}^{\rm GW}\bigg|_{\rm q = 3}= \left(\frac{3}{16} +0.926\Omega -2.146\Omega^2 +3.817\Omega^3\right)\Omega^{-1/3}
\end{equation}
The angular momentum, denoted here as $L_{\rm bin}^{\rm GW}$, is presented in Fig.~\ref{fig:LOmega} for 
$q=1$, and $q=3$.

\begin{figure}
    \centering
    \includegraphics[width=8cm]{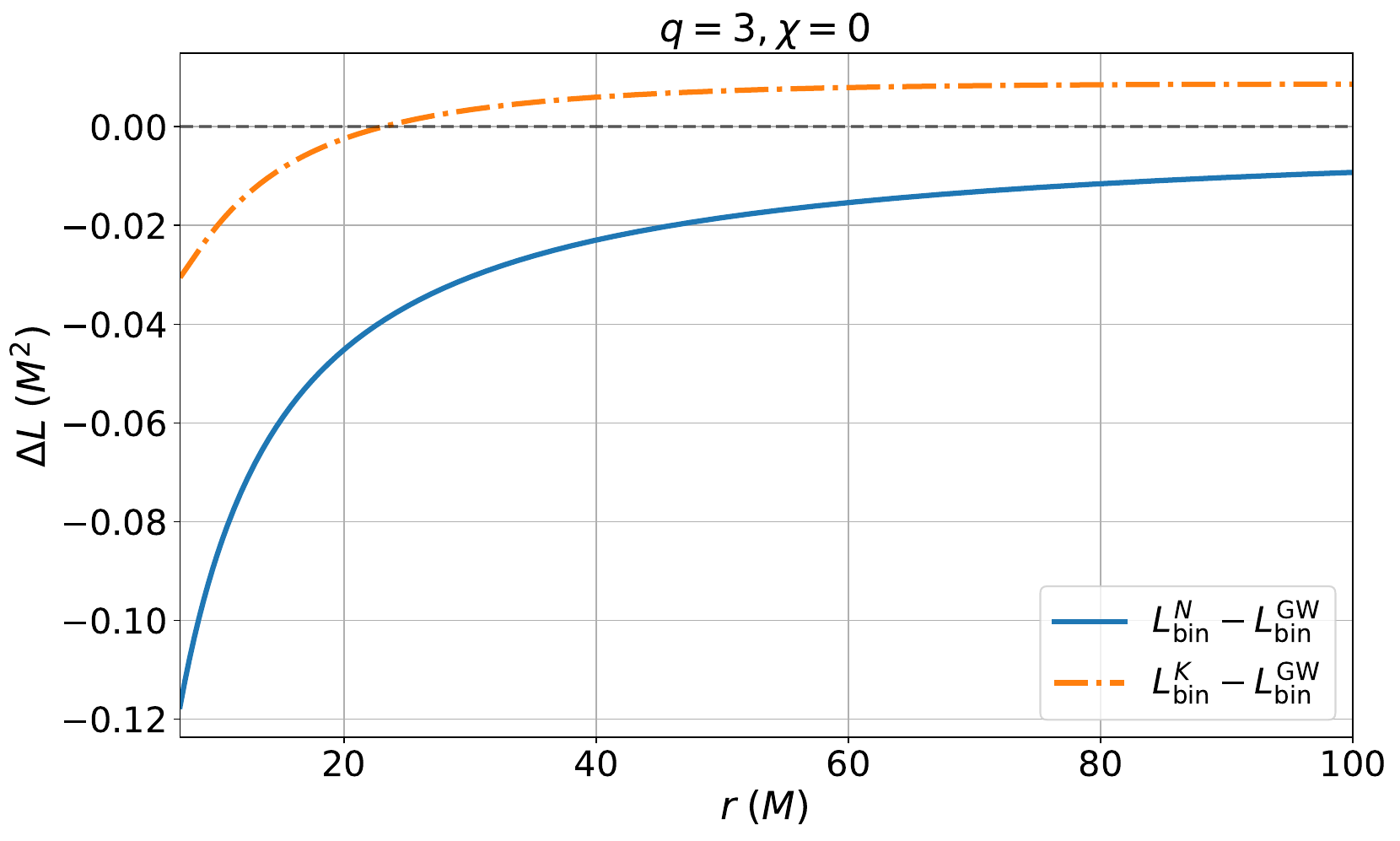}
    \caption{The cyan line represents the difference between radiated GW angular momentum flux~($L^{\rm GW}_{\rm bin}$) and the Newtonian angular momentum $L^{\rm N}_{\rm bin}$, for $q=3$ and $\chi = 0$. The difference between $L^{\rm K}_{\rm bin}$, the Kerr-inspired results, and 
    $L^{\rm GW}_{\rm bin}$, is depicted in the orange line. The dashed black line is included in the plot to serve as a reference.}
    \label{fig:delL_q3}
\end{figure}

\begin{figure}[h]
    \centering
    \includegraphics[width=8cm]{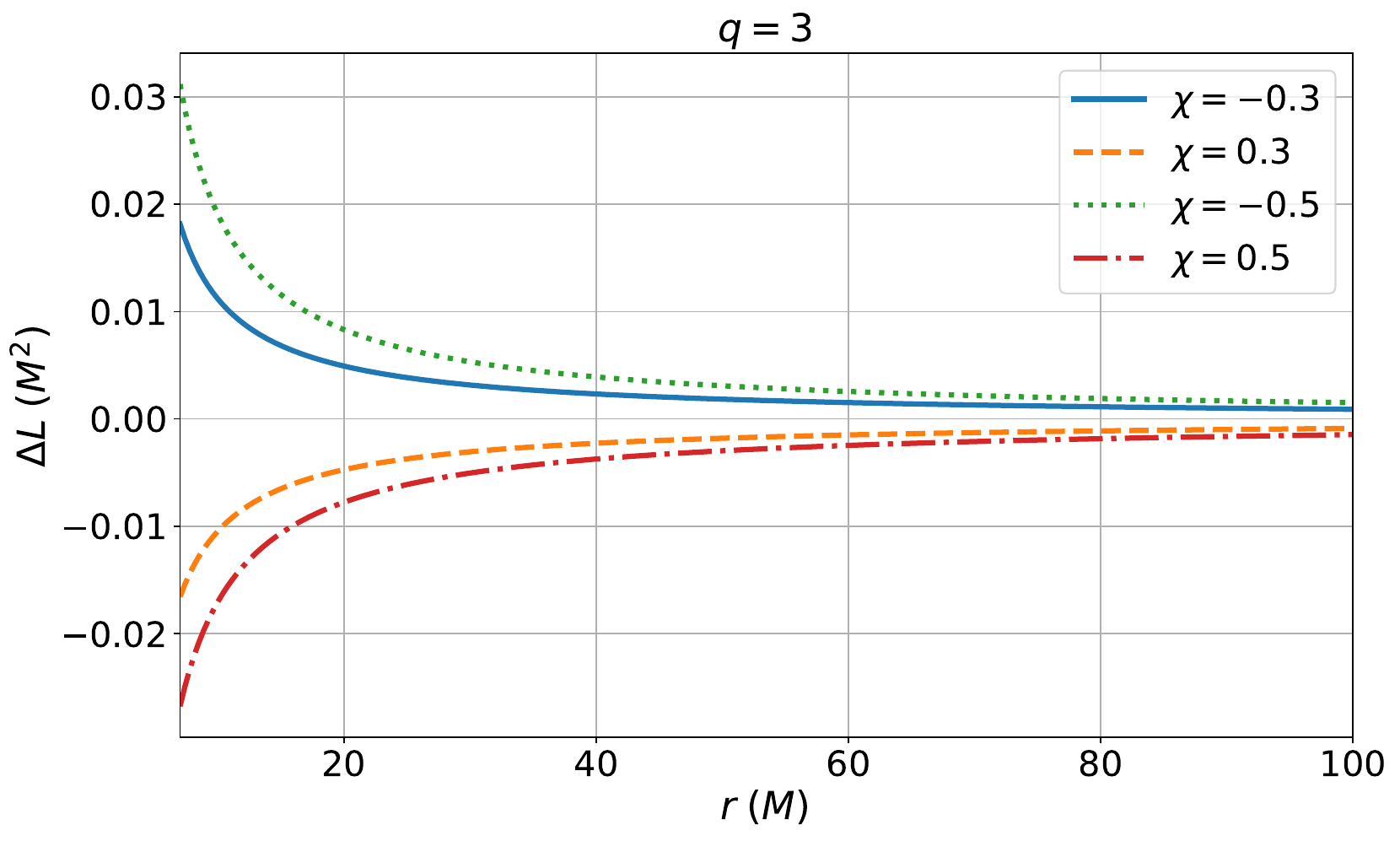}
    \caption{This plot depicts the deviation of $L^{\rm K, q=3}_{\rm bin}$ for $\chi \neq 0$ from $L^{\rm K, q=3}_{\rm bin}$ for $\chi=0$ case as the binary approaches ISCO. }
    \label{fig:delL_all_chi}
\end{figure}

Comparisons are given in Table~\ref{tab:Lbinq1nospin}.
In that table, the reasonably good agreement of $L^{\rm GW}_{\rm bin}$ and $L^{\rm K}_{\rm bin}$, as $r$ decreases, is an important result. It leads to two important conclusions. First, it adds confidence in our GW-based BOAM, the BOAM that can be used as a conserved quantity in the analysis of angular momentum exchange. Second, as $r$ decreases and GR effects become important, it agrees better with $L^{\rm K}_{\rm bin}$ than with $L^{\rm N}_{\rm bin}$. The same fact can also be observed in Fig.~\ref{fig:delL_q3}. This result for binaries without black hole spin adds confidence in applying our approach to adapting Kerr PP results to comparable mass binaries with spin. 
In Table~\ref{tab:Lbinq1spin} and Fig.~\ref{fig:delL_all_chi} we have presented our method for spinning cases.

\begin{table}[h]
\caption{Binary angular momentum $L_{\rm bin}$, in units of the square of the total mass, computed by the 
approach given in the text, as a function of the binary separation $r$ in units of the total binary mass, 
for black holes with no initial spin. Values are given for $L^{\rm K}_{\rm bin}$, the Kerr-based result given in Eq.~\eqref{eq:Lk}, 
for $L^{\rm N}_{\rm bin}$ the Newtonian value, and for  $L^{\rm GW}_{\rm bin}$, the angular momentum inferred 
from the NR data for emitted angular momentum. Results are given for $q\equiv m_1/m_2=1$ and $3$.  The transition 
from quasi-adiabatic inspiral to plunge -- the binary equivalent of the ISCO, inferred from waveforms -- is 
somewhat less than  separation of around 7$M$.} \label{tab:Lbinq1nospin}
$$
\begin{array}{|c|l|l|l||l|l|l|}\hline
  {r}  & L^{{\rm K},q=1}_{\rm bin}& L^{{\rm N},q=1}_{\rm bin}&L_{\rm bin}^{{\rm GW},q=1}& 
 L^{{\rm K},q=3}_{\rm bin} & L^{{\rm N},q=3}_{\rm bin}&L_{\rm bin}^{{\rm GW},q=3} 
    \\ \hline
  100 & 2.5190 & 2.5000 & 2.5112 & 1.8928 & 1.8750 & 1.8842 \\
   50 & 1.7949 & 1.7678 & 1.7901 & 1.3515 & 1.3258 & 1.3442 \\
   20 & 1.1625 & 1.118  & 1.1731 & 0.8812 & 0.8385 & 0.8836 \\
   10 & 0.8575 & 0.7905 & 0.8970 & 0.6593 & 0.5929 & 0.6791 \\
    7 & 0.7462 & 0.6614 & 0.8070 & 0.5829 & 0.4961&  0.6134 \\
 \hline
\end{array}
$$
\end{table}
\begin{table}[h]
\caption{Binary angular momentum $L^{\rm K,q=1}_{\rm bin}$, in units of the square of the total mass, computed by the 
approach outlined in the text, as a function of the binary separation $r$ in units of the total binary mass, 
for black holes of equal mass ($q=1$) and equal values of BH spin $\chi$. } \label{tab:Lbinq1spin}

$$
\begin{array}{|c|l|l|l|l|l|}\hline
  {r}  &\chi=0 & \chi=+0.3&\chi=-0.3& 
 \chi=+0.5& \chi=-0.5 
    \\ \hline
  100 & 2.5190 & 2.5182 & 2.5198 & 2.5176 & 2.5203 \\
   50 & 1.7949 & 1.7933 & 1.7965 & 1.7922 & 1.7976 \\
   20 & 1.1625 & 1.1583 & 1.1668 & 1.1556 & 1.1697 \\
   10 & 0.8575 & 0.8486 & 0.8669 & 0.8429 & 0.8733 \\
    7 & 0.7462 & 0.7327 & 0.7607 & 0.7242 & 0.7708 \\
 \hline
\end{array}
$$
\end{table}

\begin{table}[h]
\caption{
Binary angular momentum $L^{\rm K,q=3}_{\rm bin}$, in units of the square of the total mass, computed by the 
approach outlined in the text, as a function of the binary separation $r$ in units of the total binary mass, 
for black holes with mass ratio 3:1 ($q=3$) and equal values of BH spin $\chi$.
} \label{tab:Lbinq3spin}
$$
\begin{array}{|c|l|l|l|l|l|}\hline
  {r}  &\chi=0 & \chi=+0.3&\chi=-0.3& 
 \chi=+0.5& \chi=-0.5 
    \\ \hline
  100 & 1.8928 & 1.8920 & 1.8938 & 1.8914 & 1.8944 \\
   50 & 1.3515 & 1.3497 & 1.3533 & 1.3485 & 1.3546  \\
   20 & 0.8812 & 0.8765 & 0.8862 & 0.8735 & 0.8896  \\
   10 & 0.6593 & 0.6488 & 0.6704 & 0.6422 & 0.6783  \\
    7 & 0.5829 & 0.5664 & 0.6011 & 0.5562 & 0.6141  \\
 \hline
\end{array}
$$
\end{table}


\section{Conclusions}\label{sec:Conc}
We have provided an important element needed for a NR-based  analysis of the 
transfer of spin to orbital angular momentum, and its connection to 
black hole tidal coupling: orbital angular momentum that is meaningful for conservation of total angular momentum. This quantity is constructed by formulating a kinematic binary 
expression related to the rate of emission of gravitational wave angular momentum in the absence of spin. We have found that this NR-based binary angular momentum differs significantly from Newtonian binary angular momentum at small separation, but is in remarkably good agreement with a binary angular momentum inferred from a particle perturbation treatment rooted in the Kerr geometry.\footnote{In calculations performed subsequent to the original submission of this paper we used our BOAM prescription and looked at the change in angular momentum from the initial configuration to the ISCO, and compared it to the radiated angular momentum during that time. The agreement was very good, adding confidence in the method of Sec.~\ref{sec:KerrInspired}.}


The success of our approach to a binary angular momentum is a reason for some confidence that the related approach for angular velocity is more accurate than the Newtonian Kepler law at the small separations at which tidal transfer is significant.

The elements provided in this paper will be used in the formulation of a paper~\cite{PaperII} 
on tidal torque inferred from the waveforms in the SXS catalog.

\bigskip
{\em Acknowledgments:}
R.B and G.K. acknowledge support from NSF Grants No. PHY-2106755 and No. 
DMS-1912716. All computations were performed on the UMass-URI UNITY HPC/AI 
cluster at the Massachusetts Green High-Performance Computing Center (MGHPCC).

\pagebreak


\begin{thebibliography}{11}
\bibitem{binaryISCO} For a radiating binary of comparable masses there is no conserved 
mechanical energy, so no ``ISCO" in the PP sense. For simplicity, we shall continue to 
use the term ISCO to mean the onset of the plunge. The actual time of the plunge can 
be well approximated from the SXS waveforms.
\bibitem{initialvsref} The ``initial" parameters of a model are the data fed into the 
initial solver. The SXS catalog also provides parameters at a ``reference" time, after 
the early ``junk" radiation has dissipated. For models 0161 and 0164, the differences 
are less than 10\%.
\end{thebibliography}
\end{document}